\documentclass[12pt]{article}
\usepackage{graphicx}
\usepackage{amssymb}
\usepackage{bm}
	
\renewcommand{\abstract}[1]{{ \footnotesize \noindent {\bf Abstract} #1 \\}}
\renewcommand{\author}[1]{\subsubsection*{\it#1}}
\newcommand{\address}[1]{\subsubsection*{\it#1}}

\newcommand{\chapter}[1]{{\Large \bf \noindent #1}}

\begin{document}

\chapter{Cosmological Constraint on the Effective Number of Neutrino Species}

\author{Kazuhide Ichikawa}
\address{Institute for Cosmic Ray Research, University of Tokyo, Kashiwa 277-8582, Japan}
\abstract{
We discuss constraints on the effective number of neutrino species $N_\nu$ from recent cosmological observations such as CMB, LSS, BBN, including our own analysis which uses the WMAP and the Luminous Red Galaxy power spectrum data. We also discuss their implications on some non-standard cosmological scenarios such as the low (MeV-scale) reheating temperature scenario and the scenario with decaying particles between BBN and structure formation.}

\section{Introduction}
\label{sec:introduction}
We know very well about the energy density of relic photons from the hot Big Bang since
the COBE measurement confirmed a Planck distribution of the background photons at the temperature $T=2.725$\,K \cite{Mather:1991pc, Mather:1998gm}. In the standard cosmology, for relic radiation (relativistic) component other than photons, we assume three generations of neutrinos with a Fermi distribution, but we have not directly measured them. Also, there may be other contribution which cannot be directly detected. Such uncertainty in our knowledge of the radiation energy density in the universe is often expressed as the effective number of neutrino species $N_\nu$. $N_\nu$ is the radiation energy density (other than that of photons) normalized by the neutrino energy density for one species in the standard cosmology.\footnote{
More precisely, $N_\nu = (\rho_{\rm rel}-\rho_\gamma)/\rho_{\nu,{\rm thm}}$, where $\rho_\gamma$ is the photon energy density, $\rho_{\rm rel}$ is the total energy density of photons, three active species of neutrinos and extra relativistic contribution, and $\rho_{\nu,{\rm thm}}$ is defined as  $\rho_{\nu,{\rm thm}}=(7\pi^2/120) (4/11)^{4/3} T ^4$ using the photon temperature $T$ after the electron-positron annihilation. $\rho_{\nu,{\rm thm}}$ corresponds to the energy density of a single species of neutrino, assuming that neutrinos are completely decoupled from the electromagnetic plasma before the electron-positron annihilation takes place and they obey a Fermi-Dirac distribution.}
We can think of many candidates in particle physics/cosmology that could modify the standard value of $N_\nu = 3$. There may be sterile neutrinos, gravitational waves, axions, majorons, large lepton asymmetries, MeV-scale reheating and so on. Therefore it is of great importance to constrain $N_\nu$ from the cosmological observations to probe a certain class of models in the particle physics/cosmology.

Recent precise observations of the cosmic microwave background (CMB) anisotropies and large scale structure (LSS) make it
possible to measure $N_\nu$ through its effects on the growth of cosmological perturbations. 
These effects come from the fact that the density perturbation does not 
grow (the gravitational potential decays) during the radiation-dominated era. 
Specifically, a more relativistic degree of freedom causes greater 
early integrated Sachs-Wolfe effect on the CMB power spectrum,
which leads to higher first peak height.
Also, since it delays the epoch of the matter-radiation equality and makes the horizon
 at that time larger, the turnover position of the matter power
spectrum is shifted to larger scales and the power at smaller scales is
suppressed. Therefore, by observing the CMB and LSS, we can measure $N_\nu$ during structure formation. 
In detail, assuming the smallest scale relevant to our observations to be about 5\,Mpc,
since structure formation of that scale begins around the temperature $T\approx 20$\,eV
 (at which the scale enters the horizon), these observations probe $N_\nu$ at $T \lesssim 20$\,eV (if 1\,Mpc is taken to be the smallest observable scale, as is the case with Lyman-$\alpha$ forest measurements (Ly$\alpha$) , it would be $T \lesssim 100$\,eV). Thus, the CMB and LSS can measure $N_\nu$ independently of another well-known probe, big bang nucleosynthesis (BBN), which measures 
it in much earlier universe around $T=O({\rm MeV})$.

There are now many works to constrain $N_\nu$ from cosmological observations and, surprisingly, some data sets seem to favor $N_\nu > 3$ as found in Refs.~\cite{Spergel:2006hy,Seljak:2006bg,Cirelli:2006kt,Mangano:2006ur} (we show the result of Ref.~\cite{Seljak:2006bg} in Table \ref{tab:Nnucomparison}). The galaxy power spectrum of 2dF \cite{Cole:2005sx} favors standard $N_\nu = 3$ but the SDSS main galaxy power spectrum \cite{Tegmark:2003uf} and the Ly$\alpha$ data prefer $N_\nu > 3$. Thus, we would like to cross-check these disagreements by using the SDSS luminous red galaxy (LRG) power spectrum \cite{Tegmark:2006az} which has more statistical constraining power than 2dF or SDSS main sample (the effective volume of the LRG survey is about 6 times larger than that of the SDSS main galaxy sample and over 10 times larger than that of the 2dFGRS \cite{Tegmark:2006az}).\footnote{Recently, Ref.~\cite{Hamann:2007pi} argues that the discrepancies are due to systematic effects of scale-dependent biasing in the galaxy power spectrum.}
 
We describe our analysis and result in the next section. The implications for some non-standard cosmological scenarios are discussed in Sec.~\ref{sec:implication}. We summarize in Sec.~\ref{sec:summary}.
 
\begin{table}
\begin{center}
\begin{tabular}{|l|l|l|}
\hline
  & 95\% limit & Data set \\
  \hline
Seljak, Slosar, McDonald \cite{Seljak:2006bg} & $N_\nu = 5.3^{+2.1}_{-1.7} $ & All \\
 & $N_\nu = 4.8^{+1.6}_{-1.4}$ & All + HST \\
 & $N_\nu = 6.0^{+2.9}_{-2.4}$ & All $-$ BAO \\
 & $N_\nu = 3.9^{+2.1}_{-1.7}$ & All $-$ Ly$\alpha$ \\
 & $N_\nu = 7.8^{+2.3}_{-3.2}$ & WMAP3+SN+SDSS(main) \\
 & $N_\nu = 3.2^{+3.6}_{-2.3}$ & WMAP3+SN+2dF\\
 & $N_\nu = 5.2^{+2.1}_{-1.8}$ & All-2dF-SDSS(main) \\
\hline
Ichikawa, Kawasaki, Takahashi \cite{Ichikawa:2006vm}&  $N_\nu = 3.1^{+5.1}_{-2.2}$ & WMAP3+SDSS(LRG) \\
\hline
\end{tabular}
\end{center}
\caption{Comparison of $N_\nu$ constraints using various data set combinations. ``All" refers to WMAP3 + other CMB + Ly$\alpha$ + galaxy power spectrum (SDSS main sample + 2dF) + SDSS baryon acoustic oscillation (BAO) + Supernovae Ia (SN). See Ref.~\cite{Seljak:2006bg} for details. SDSS (main) and Ly$\alpha$ favor $N_\nu > 3$.}
\label{tab:Nnucomparison}
\end{table}

\section{Constraint on $N_\nu$ from WMAP and the SDSS LRG power spectrum}
\label{sec:constraint}
We summarize the analysis and result obtained in Ref.~\cite{Ichikawa:2006vm}.
We constrain $N_\nu$ in the flat $\Lambda$CDM universe with the initial perturbation power spectrum which is adiabatic and described by a power law. This model has six cosmological parameters: the baryon density $\omega_b$, the matter density $\omega_m$, the normalized Hubble constant $h$, the reionization optical depth $\tau$, the scalar spectral index of primordial perturbation power spectrum $n_s$ and its amplitude $A$ ($\omega = \Omega h^2$, where $\Omega$ is the energy density normalized by the critical density). Theoretical CMB and matter power spectra are calculated by the CMBFAST code \cite{Seljak:1996is} and $\chi^2$ by the likelihood codes of the WMAP three-year data \cite{Jarosik:2006ib,Hinshaw:2006ia,Page:2006hz} and of the SDSS LRG power spectrum data \cite{Tegmark:2006az}. We apply modelling of nonlinearity and scale-dependent bias as in Ref.~\cite{Tegmark:2006az} to the linear matter power spectrum before fitting to the LRG data. This modeling has two parameters, the galaxy bias factor $b$ and nonlinear correction factor $Q_{\rm nl}$. Specifically, we connect the linear matter power spectrum $P_{\rm lin}(k)$ and the galaxy power spectrum $P_{\rm gal}(k)$ by $P_{\rm gal}(k) = b^2\{ (1+Q_{\rm nl}\, k^2)/(1+ 1.4\, k)\} P_{\rm lin}(k)$.
 We calculate the $\chi^2$ as functions of $N_\nu$ by marginalizing over the above parameters (6 parameters for WMAP alone and 8 for WMAP+SDSS). The marginalization is carried out by the Brent minimization \cite{brent} modified to be applicable to multi-dimension parameter space as described in Ref.~\cite{Ichikawa:2004zi}. 

\begin{figure}
\begin{center}
\includegraphics[scale=0.8]{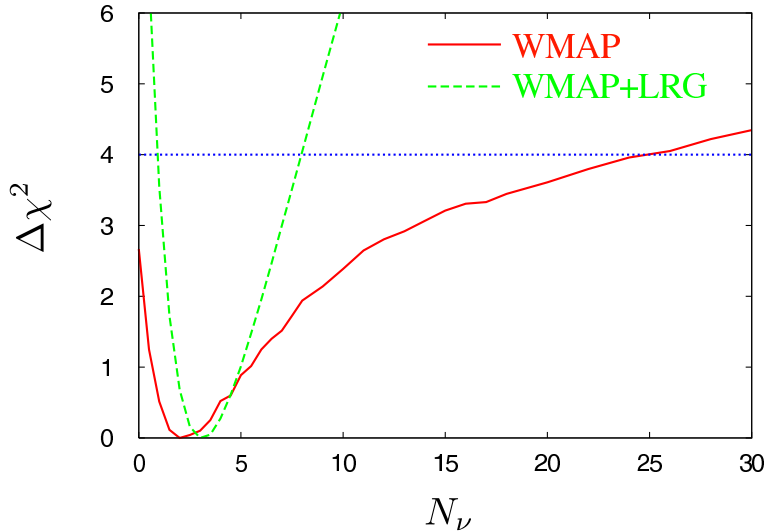} 
\end{center}
\caption{$\Delta \chi^2$ as functions of $N_\nu$. The red solid line shows the result of using the WMAP three-year data alone and the green dashed line that of using the WMAP three-year data and the SDSS LRG power spectrum \cite{Ichikawa:2006vm}.}
\label{fig:chi2}
\end{figure}

 We show the results of $\chi^2$ minimization in Fig.~\ref{fig:chi2}. We have checked that the results for the standard three neutrino species agree with the WMAP \cite{Spergel:2006hy} and SDSS \cite{Tegmark:2006az} groups' analyses. For the WMAP three-year alone case, it has been checked in Ref.~\cite{Fukugita:2006rm} that the best fit $\chi^2$ and parameters agree. With regard to the WMAP and LRG combined analysis, our best fit parameter values for the three neutrino species are $\omega_b=0.0222 \pm 0.0007$, $\omega_m=0.1288 \pm 0.0044$, $h=0.718 \pm 0.018$, $\tau = 0.088 \pm 0.029$, $n_s=0.958 \pm 0.016$, $\sigma_8 = 0.770 \pm 0.033$ (we report here $\sigma_8$ instead of $A$ to compare with Ref.~\cite{Tegmark:2006az}), $b=1.877 \pm 0.065$ and $Q_{\rm nl}=30.4 \pm 3.5$. The central values fall well within the 1$\sigma$ ranges of the constraints derived in Ref.~\cite{Tegmark:2006az} and the 1$\sigma$ errors are almost identical to those quoted in Ref.~\cite{Tegmark:2006az}.
 
 The limits corresponding to $\Delta \chi^2 = 4$ are $N_\nu < 25$ for WMAP three-year data alone and $0.8 < N_\nu < 8.0$ for WMAP and SDSS LRG combined. Since the $\chi^2$ functions show some asymmetric features, we derive 95\% confidence limits by integrating the likelihood functions ${\cal L} = \exp({-\Delta \chi^2/2})$. This yields a 95\% C.L. bound of $N_\nu < 42$ for WMAP alone and $0.9 < N_\nu < 8.2$ for WMAP+LRG.

Our constraint on $N_\nu$ from the WMAP three-year CMB power spectrum and the SDSS LRG power spectrum, can be summarized as 
\begin{eqnarray}
0.9 < N_\nu < 8.2 \quad {\rm or} \quad N_\nu = 3.1^{+5.1}_{-2.2} \label{eq:Nnu_constraint}
\end{eqnarray}
at the 95\% C.L. The minimum $\chi^2$ is given at the value quite close to the standard model value but the upper bound is not so stringent that it is compatible with the Ly$\alpha$'s high value. Since it is very difficult to remove systematic errors from the galaxy clustering and Ly$\alpha$ analyses, we may have to wait for next generation CMB experiments to significantly improve our knowledge on $N_\nu$ (the PLANCK sensitivity for $N_\nu$ is forecasted to be $\sim 0.2$; see $e.g.$ Ref.~\cite{Ichikawa:2006dt}).
 
\section{Implications for non-standard cosmology}
\label{sec:implication}
\subsection{Low reheating temperature}
 In this scenario, the effective number of neutrinos $N_\nu$ can be less than the standard value since
 neutrinos are not thermalized.
We briefly explain this scenario and how $N_\nu$ and the reheating temperature $T_R$ are related. 
For more details, we refer to Ref.~\cite{Ichikawa:2005vw}.

The standard big bang model assumes that the universe
was once dominated by thermal radiation composed of
photons, electrons, neutrinos, and their antiparticles.
The reheating temperature is the temperature at which the universe becomes such a radiation-dominated state
and it is usually assumed to be so high that every particle species is in thermal equilibrium. 
In particular, neutrinos are considered to obey a Fermi distribution. 
But if the reheating temperature is as low as a few MeV, their distribution function deviates from the thermal one.
In contrast to electrons that are always (at least until the
temperature drops below a few eV) in thermal contact
with photons via electromagnetic forces, neutrinos interact
with electrons and themselves only through the weak interaction.
The decoupling temperature of the neutrinos
should be around 3 MeV for the electron neutrinos and
5 MeV for the muon and tau neutrinos, respectively (the difference comes from the fact that the electron
neutrinos have additional charged current interaction with
electrons). Therefore the neutrinos might not be fully thermalized and lead to $N_\nu < 3$
if the reheating temperature is in the MeV range. In fact, such a low reheating temperature can be found in many cosmological scenarios
especially in order to dilute the unwanted relics such as
the gravitinos.

\begin{figure}
\begin{center}
\includegraphics{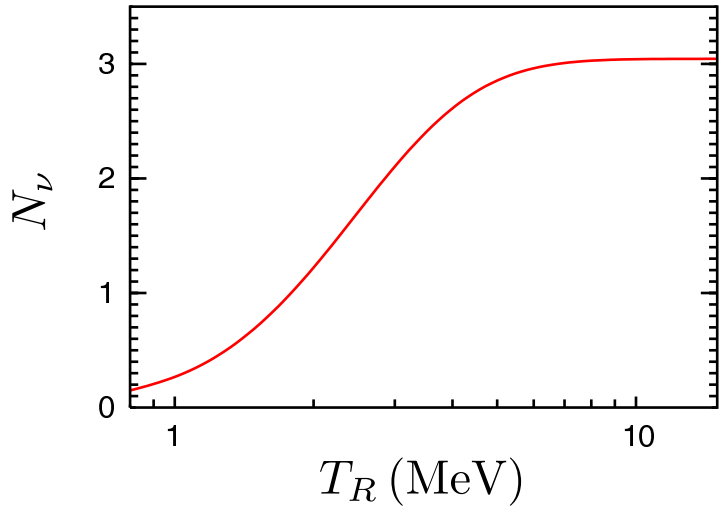} 
\end{center}
\caption{The relation between the effective neutrino number $N_\nu$ and the reheating temperature $T_R$ \cite{Ichikawa:2005vw}.}
\label{fig:TR_Nnu}
\end{figure}

In Ref.~\cite{Ichikawa:2005vw}, we have calculated how much neutrinos are thermalized 
when $T_R = O({\rm MeV})$ and have derived the relation between $T_R$ and $N_\nu$
which is shown in Fig.~\ref{fig:TR_Nnu}.\footnote{
As discovered in Ref.~\cite{Ichikawa:2005vw}, it should be noted that the $^4$He abundance, $Y_p$, increases while $N_\nu$ decreases in this scenario. This is in contrast to the conventional non-standard $N_\nu$ scenario where decreasing $N_\nu$ accompanies decreasing $Y_p$. The difference occurs because non-thermal distribution of (electron-type) neutrinos modifies the neutron-proton conversion rate, which significantly affects $Y_p$.}
We can convert our constraint on $N_\nu$, Eq.~(\ref{eq:Nnu_constraint}), into the lower bound 
on $T_R$ using Fig.~\ref{fig:TR_Nnu}:
\begin{eqnarray}
T_R > 2\, {\rm MeV}.
\end{eqnarray}

\subsection{Decaying particles between BBN and structure formation}
As mentioned at the end of Sec.~\ref{sec:constraint}, our analysis using the SDSS LRG power spectrum cannot exclude $N_\nu = 4 \sim 5$ which is favored by the Ly$\alpha$ data. On the other hand, the BBN constraint from the recent measurements of the $^4$He abundance, $Y_p$, seems to favor $N_\nu \sim 3$ as we summarize in Table \ref{tab:NnuBBN}. If these Ly$\alpha$ and $Y_p$ analyses are correct, can we obtain a consistent cosmological scenario ? 

Recall that they measure $N_\nu$ at different cosmological epochs as noted in Sec.~\ref{sec:introduction}. Thus, we can do this by increasing $N_\nu$ after BBN but before structure formation begins. This corresponds to the temperature between about 1\,MeV and 100\,eV or the time between 1\,s and $10^8$\,s. In Ref.~\cite{Ichikawa:2007jv}, we have investigated models with decaying particles which have a lifetime of $O(1$--$10^8)$\,s and can contribute to the required additional $N_\nu$. We have considered (a) saxion decay into two axions; (b) gravitino
decay into axino and axion; (c) Dirac right-handed sneutrino decay
into gravitino and right-handed neutrino, and have shown that there are parameter regions which can make $\Delta N_\nu \sim 1$ and do not affect other observations. 

\begin{table}
\begin{center}
\begin{tabular}{|l|l|l|}
\hline
   & $Y_p$ ($1\sigma$) & $N_\nu$ (95\% limit)  \\
  \hline
Olive, Skillman \cite{Olive:2004kq} & $0.249 \pm 0.009$ & $3.1^{+1.4}_{-1.2} $  \\
Fukugita, Kawasaki \cite{Fukugita:2006xy} & $0.250 \pm 0.004$ &  $3.20^{+0.76}_{-0.68}$  \\
Peimbert, Luridiana, Peimbert \cite{Peimbert:2007vm} & $0.2427 \pm 0.0028$ &  $3.01^{+0.52}_{-0.48}$  \\
Izotov, Thuan, Stasinska \cite{Izotov:2007ed} & $0.2516 \pm 0.0011$ &  $3.32^{+0.23}_{-0.24}$  \\
\hline
\end{tabular}
\end{center}
\caption{Comparison of $N_\nu$ constraints from recent $Y_p$ measurements. We also used the observed deuterium abundance ${\rm D/H} = (2.82 \pm 0.27) \times 10^{-5}$ \cite{O'Meara:2006mj} and the BBN fitting formula in Ref.~\cite{Serpico:2004gx}. $N_\nu > 4$ is not favored by the three recent measurements.}
\label{tab:NnuBBN}
\end{table}

\begin{figure}
\begin{center}
\includegraphics{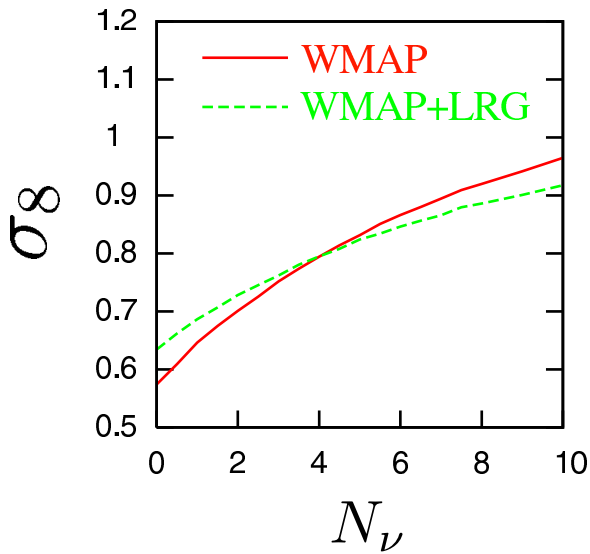} 
\end{center}
\caption{The values of $\sigma_8$ when we minimize $\chi^2$ for the WMAP three-year data alone (the red solid line) or together with the SDSS LRG power spectrum (the green dashed line).}
\label{fig:Nnu_sigma8}
\end{figure}

Finally, we would like to comment that the reason why the Ly$\alpha$ data favor $N_\nu > 3$ is connected to the $\sigma_8$ value measured by the Ly$\alpha$ which is higher than the WMAP3 value, as is pointed out by Ref.~\cite{Seljak:2006bg}. In the standard $\Lambda$CDM model ($i.e.$ $N_\nu=3$), WMAP3 alone gives $\sigma_8 = 0.76 \pm 0.05$ \cite{Spergel:2006hy} while combining it with the Ly$\alpha$ gives much higher value $\sigma_8 = 0.85 \pm 0.02$ \cite{Seljak:2006bg}. However, if we remove the assumption $N_\nu = 3$ and increase $N_\nu$, the best fit value of $\sigma_8$ from WMAP increases (we can see this explicitly in Fig.~\ref{fig:Nnu_sigma8}) and more agrees with the Ly$\alpha$ data. 

Now, turning the argument around, we can say that observations which give relatively higher $\sigma_8$ basically favor $N_\nu > 3$. We know that some of the weak lensing observations indicate high $\sigma_8$. For example, Ref.~\cite{Benjamin:2007ys} gives $\sigma_8 = 0.84 \pm 0.07$ using a ground-based telescope and Ref.~\cite{Massey:2007gh} gives $\sigma_8 = 0.95^{+0.093}_{-0.075}$ using a space telescope. Ref.~\cite{Lesgourgues:2007te} combined the latter with the Ly$\alpha$ data and obtained, naturally, a high value $\sigma_8 = 0.87 \pm 0.05$. Although the statistical significance is not very high, the weak lensing data too give motivation for the scenario presented in this section.

\section{Summary}
\label{sec:summary}
We have discussed the cosmological constraint on the effective neutrino species $N_\nu$ and its implications for some of 
non-standard cosmological scenarios. Our analysis showed that the constraint from the CMB data of WMAP3 combined with the galaxy power spectrum of the SDSS LRG sample is $0.9 < N_\nu < 8.2$ at 95\% C.L. The lower bound on $N_\nu$ can be converted to the lower bound on the reheating temperature, $T_R > 2$\,MeV. The upper bound turned out to be not very stringent so the preference of $N_\nu > 3$ indicated in the Ly$\alpha$ data cannot be ruled out. This motivates us to investigate a scenario in which $N_\nu = 3$ during BBN (as constrained from the $^4$He abundance) and $N_\nu > 3$ before structure formation. $N_\nu > 3$ is considered to be also favored by some of the weak lensing observations which give high $\sigma_8$ values as the Ly$\alpha$ data. We found that we can increase $N_\nu$ by decaying particles within the framework of supersymmetric extensions of standard model of particle physics. 


\end{document}